\documentclass[showpacs,twocolumn,aps,preprintnumbers,letterpaper]{revtex4}
\usepackage{amsmath,amssymb}
\usepackage{epsfig}
\usepackage{graphicx}
\usepackage{amsmath}
\usepackage{slashed}
\usepackage{amsfonts}
\usepackage{epstopdf}

\newcommand{\be}{\begin{equation}}
\newcommand{\ee}{\end{equation}}
\newcommand{\bear}{\begin{eqnarray}}
\newcommand{\eear}{\end{eqnarray}}
\newcommand{\ba}{\begin{array}}
\newcommand{\ea}{\end{array}}

\def\be{\begin{eqnarray}}
\def\ee{\end{eqnarray}}
\def\bea{\be}
\def\eea{\ee}

\newcommand{\e}{{\mbox{e}}}

\def\roughly#1{\mathrel{\raise.3ex\hbox{$#1$\kern-.75em%
\lower1ex\hbox{$\sim$}}}}

\begin{document}

\title{ The Instanton-Dyon Liquid Model V: \\Twisted Light Quarks }

\author{Yizhuang Liu}
\email{yizhuang.liu@stonybrook.edu}
\affiliation{Department of Physics and Astronomy, Stony Brook University, Stony Brook, New York 11794-3800, USA}

\author{Edward Shuryak}
\email{edward.shuryak@stonybrook.edu}
\affiliation{Department of Physics and Astronomy, Stony Brook University, Stony Brook, New York 11794-3800, USA}

\author{Ismail Zahed}
\email{ismail.zahed@stonybrook.edu}
\affiliation{Department of Physics and Astronomy, Stony Brook University, Stony Brook, New York 11794-3800, USA}


\date{\today}
\begin{abstract}
We discuss an extension of the instanton-dyon liquid model that includes twisted light quarks in the fundamental representation 
with explicit $Z_{N_c}$ symmetry for the case with equal number of colors $N_c$ and flavors $N_f$. We map the model  on a 3-dimensional quantum effective theory, 
and analyze it in the mean-field approximation. The effective potential 
and the vacuum chiral condensates are made explicit for $N_f=N_c=2, 3$. The low temperature phase is center symmetric 
but breaks spontaneously flavor symmetry with  $N_f-1$ massless pions. The high temperature phase breaks center symmetry but supports finite  and unequal quark condensates.
\end{abstract}


\maketitle

\setcounter{footnote}{0}


\section{Introduction}

In the QCD ground state confinement and chiral symmetry breaking are intertwined 
as lattice simulations have now established~\cite{LATTICE}. The loss of confinement 
with increasing temperature as described by a jump in the Polyakov line is followed
by a rapid cross-over in the chiral condensate for $2+1$ flavors. When the quarks are
in the adjoint representation, the cross over occurs much later than the deconfinement
transition. There is increasing lattice evidence that the topological nature of the underlying
gauge configurations maybe key in understanding some aspects of these results~\cite{CALO-LATTICE}.

This work is a continuation of our earlier studies~\cite{LIU1,LIU2,LIU3,LIU5}  of the  gauge 
topology using the instanton-dyon liquid model. The starting point of the model are the KvBLL
instantons  threaded by finite holonomies and their splitting into instanton-dyon constituents
 \cite{KVLL}, with strong semi-classical interactions~\cite{DP,DPX,LARSEN}.  At low temperature,  
 the phase preserves center symmetry but breaks 
 spontaneously chiral symmetry. At sufficiently high temperature, the phase restores  both 
 symmetries as the  constituent instanton-dyons regroup into topologically neutral 
 instanton-anti-instanton molecules. The importance of  fractional topological constituents for
 confinement was initially suggested through instanton-quarks in~\cite{ARIEL}, and more recently
 using  bions in~\cite{UNSAL}.

The instanton-dyons carry fractional topological charge $1/N_c$ and are able to localize chiral
quarks into zero modes. For  quarks in the fundamental representation, as the KvBLL fractionate,
the zero-mode migrates to the heavier instanton-dyon constituent~\cite{KRAAN}.  The random 
hopping of these zero modes in the instanton-dyon liquid is at the origin of the spontaneous
breaking of chiral symmetry as has  been shown both numerically~\cite{SHURYAK1,SHURYAK2}
and using mean-field methods~\cite{LIU2}. In supersymmetric QCD some arguments were 
presented in~\cite{TIN}.

At finite temperature the light quarks are subject to anti-periodic boundary conditions on $S^1$ 
to develop the correct occupation statistics in bulk. General twisted fermionic boundary conditions 
on $S^1$ amounts to thermal QCD with Bohm-Aharanov phases that alter fundamentally the nature
of  the light quarks~\cite{RW,RMT}. A particularly interesting proposal consists of a class of $Z_{N_c}$
 twisted QCD boundary conditions with $N_c=N_f$ resulting in a manifestly $Z_{N_c}$ symmetric 
QCD dubbed $Z_{N_c}$-QCD~\cite{JAP}. The confined phase is both center and chiral symmetric
eventhough the boundary conditions are flavor breaking. The deconfined phase is center and chiral
symmetry broken~\cite{JAP,TAKUMI}. 

The purpose of this paper is to address some aspects of twisted fermionic boundary conditions
in the context of the instanton-dyon liquid model. Since the localization of the
zero-modes on a given instanton species is very sensitive to the nature of the the twist on $S^1$, 
this deformation offers an insightful tool for the possible understanding of the fundamental aspects
of the spontaneous breaking of chiral symmetry through the underlying topological constituents.  
Similar issues were addressed using PNJL models~\cite{JAP} and more recently monopole-dyons
and without anti-monopole-dyons for small $S^1$~\cite{THOMAS}.  A numerical analysis 
in the the instanton-dyon liquid model with $N_f=N_c=2$  was recently  presented in~\cite{LARS3}.

In section 2 we briefly review the model and discuss the general case of $N_c=N_f$ twisted boundary conditions.
The special cases of $N_c=N_f=2,3$ are given and the corresponding normalizable zero-modes around the center
symmetric point constructed. We derive explicitly the pertinent hopping matrices between the instanton-dyons and
the instanton-anti-dyons for the case of $N_c=N_f=2,3$ which are central to the quantitative study of the 
spontaneous breaking of chiral symmetry.
In section 3 we use a series of fermionization and bosonization transformations to map the instanton-dyon partition function
on a 3-dimensional effective theory. For $N_f>2$, 
additional  discrete symmetries combining charge conjugation and exchange between conjugate flavor pairs
are identified, with the same chiral condensates at high temperature. In section 4 we derive the effective
potential for the ground state of the 3-dimensional effective theory. We explicitly show that it supports a center symmetric state with spontaneously broken chiral symmetry. The center asymmetric phase at high temperature 
supports unequal chiral condensates. Our conclusions are in section 5.

\section{ Effective action with twisted fermions}

\subsection{General setting}

For simplicity we detail here the general setting for $N_c=2$.
The pertinent changes for any $N_c$ will be quoted when 
appropriate. For a fixed holonomy with $A_4(\infty)/2\omega_0=\nu \tau^3/2$ and  $\omega_0=\pi T$, the
SU(2) KvBLL instanton~\cite{KVLL} is composed of a pair of instanton-dyons labeled by L, M 
(instanton-anti-dyons by $\overline {\rm L},\overline {\rm M}$). In general, there are are $N_c-1$ BPS
instanton-dyons  and only one twisted  instanton-dyon. As a result the global gauge 
symmetry is reduced through $SU(N_c)\rightarrow U(1)^{N_c-1}$. 

For example, the  grand-partition function   for dissociated $N_c=2$ KvBLL instantons and anti-instantons 
and $N_f$ massless flavors is

\bea
{\cal Z}_{1}[T]&&\equiv \sum_{[K]}\prod_{i_L=1}^{K_L} \prod_{i_M=1}^{K_M} \prod_{i_{\bar L}=1}^{K_{\bar L}} \prod_{i_{\bar M}=1}^{K_{\bar M}}\nonumber\\
&&\times \int\,\frac{f_Ld^3x_{Li_L}}{K_L!}\frac{f_Md^3x_{Mi_M}}{K_M!}
\frac{f_Ld^3y_{{\bar L}i_{\bar L}}}{K_{\bar L}!}\frac{f_Md^3y_{{\bar M}i_{\bar M}}}{K_{\bar M}!}\nonumber\\
&&\times {\rm det}(G[x])\,{\rm det}(G[y])\,\left|{\rm det}\,\tilde{\bf T}(x,y)\right|^{N_f}e^{-V_{D\overline D}(x-y)}\nonumber\\
\label{SU2}
\eea
Here $x_{mi}$ and $y_{nj}$ are the 3-dimensional coordinate of the i-dyon of  m-kind
and j-anti-dyon of n-kind. Here
$G[x]$ a $(K_L+K_M)^2$ matrix and $G[y]$ a $(K_{\bar L}+K_{\bar M})^2$ matrix whose explicit form are given in~\cite{DP,DPX}.
$V_{D\bar D}$ is the streamline interaction between ${\rm D=L,M}$ dyons and ${\rm \bar D=\bar L, \bar M}$ antidyons as numerically discussed in~\cite{LARSEN}. For the SU(2) case  it is Coulombic asymptotically with a core at short distances~\cite{LIU1}.
We will follow our original discussion with light quarks in~\cite{LIU2}, with the determinantal interactions in (\ref{SU2})
providing for an effective core repulsion on average.
We omit the explicit repulsion between the cores as in~\cite{LIU5}, for simplicity. 
The fugacities $f_{i}$ are related to the overall instanton-dyon density, and can be estimated using  lattice simulations~\cite{CALO-LATTICE}. Here they are external parameters, with a dimensionless density

\be
\label{NTT}
{\bf n}=\frac{4\pi \sqrt{f_Lf_M}}{\omega_0^2}\approx {\bf C}e^{-\frac{S(T)}2}
\ee
For definiteness, the KvBLL instanton  action to one-loop is

\be
{S(T)}\equiv \frac {2\pi}{\alpha_s(T)}=\left(11\frac{N_c}3-2\frac{N_f}3\right){\rm ln}\left(\frac T{0.36T_D}\right)
\ee
The fermionic  determinant ${\rm det}\,\tilde{\bf T}(x,y)$  with twisted quarks will be detailed below. 
In many ways (\ref{SU2}) resembles the partition function for the instanton-anti-instanton ensemble~\cite{ALL}.

\subsection{Twisted boundary conditions and normalizable zero modes}

Consider $N_f=N_c$ QCD on $S^1\times R^3$ with the following anti-periodic boundary conditions
modulo a flavor twist in the center of $SU(N_c)$

\be
\label{BOUND1}
\psi_f(\beta, \vec x) =-z^{f-1}\psi_f(0, \vec x) 
\ee
with $z=e^{i 2\pi/N_c}$  and $f=1,2,3, ...=u,d,s, ...$ respectively. Under a $Z_{N_c}$  twisted gauge transformation  of the type

\be
\Omega(\beta, \vec x)=z^k\Omega(0, \vec x)
\ee
(\ref{BOUND1}) is $Z_{N_C+N_f}$ symmetric following the flavour relabeling $f+k\rightarrow f$. 
As a result the theory is usually referred to as $Z_{N_c}$-QCD~\cite{JAP}. 
In contrast,  (\ref{BOUND1})  breaks explicitly chiral flavor symmetry through

\be
\label{SYMNF}
U_L(N_f)\times U_R(N_f)\rightarrow U_L^{N_f}(1)\times U_R^{N_f}(1)
\ee

To construct explicitly the  fermionic zero modes in a BPS or KK dyon with the twisted boundary
conditions (\ref{BOUND1}), we consider the generic boundary condition

\be
\label{BOUND2}
\psi(x_4+\beta,\vec x)=-e^{i\phi}\psi(x_4,\vec x)
\ee
and redefine the quark field through $\psi=e^{iT\phi x_4}\tilde \psi$. The latter satisfies
a modified Dirac equation with an imaginary chemical potential $-\phi\,T$~\cite{RW},

\be
\label{DIRAC1}
(i\gamma \cdot D-\gamma_4T\phi)\tilde \psi=0
\ee
In a BPS dyon with periodic boundary conditions, the solution to (\ref{DIRAC1}) asymptote

\be
\label{DIRAC2}
\tilde\psi\rightarrow e^{-\pi T \nu r\pm \phi T r}
\ee
which is normalizable for $|\phi|<\pi  \nu$. For anti-periodic boundary condition, the requirement
for the existence of a normalizable zero mode in a BPS dyon is $|\phi-\pi|<\pi \nu$.

\subsection{Case: $N_c=N_f=3$}

For $N_c=N_f=3$,  the flavor twisted boundary condition (\ref{BOUND1}) takes the explicit form

\bea
\label{BNC}
&&u(\beta)=-u(0)\nonumber\\
&&d(\beta)=e^{-i\pi/3}d(0)\nonumber\\
&&s(\beta)=e^{+i\pi/3}s(0)
\eea
The d,s boundary conditions in (\ref{BNC}) 
admit a discrete symmetry under the combined charge conjugation and the flavor
exchange $d\leftrightarrow s$.

The normalizability condition for the quark zero modes following from the flavor twisted boundary conditions
in (\ref{DIRAC1}-\ref{DIRAC2}) shows that  $f=1=u$ always support a normalizable KK zero mode,
while $f=2,3=d,s$ support BPS zero modes that are at the edge of the normalizability domain in
the symmetric phase with $\nu=1/3$. The BPS modes carry a time dependence of the form $e^{\pm \frac{i\omega_0}{3}x_4}$
 as $\nu \rightarrow 1/3$, while the KK mode  carries a time dependence of the form $e^{i\omega_0 x_4}$. In both cases,
 we are restricting the modes to the lowest frequencies in Euclidean $x_4$-time, for simplicity. This means a moderatly large temperature ranging from the center symmetric to asymmetric phase. 

The explicit form of the twisted zero modes in a BPS dyon and satisfying the twisted boundary condition (\ref{BOUND2}) can be
obtained in closed form in the hedgehog gauge,

\bea
\label{ZERO1}
\tilde  \psi_{\mp, A\alpha}(r)=(\alpha_1(r)\epsilon+\alpha_2(r) \sigma \cdot \hat r\epsilon)_{A\alpha}
\eea
in color-spin, with $\epsilon_{A\alpha}=-\epsilon_{\alpha A}$ and

\bea
\alpha_{1,2}(r)=&&\frac{\chi_{1,2}(r)}{\sqrt{2\pi \nu T r \sinh(2\pi \nu T r)}}\nonumber\\
\chi_1(r)=&&-\frac{\tilde\phi }{\pi  \nu}\sinh(\tilde\phi T r)+\tanh(\pi T \nu r)\cosh(\tilde\phi T r)\nonumber\\
\chi_2(r)=&&\mp\left(\frac{\tilde\phi }{\pi  \nu}\cosh(\tilde\phi T r)-\coth(\pi T \nu r)\sinh(\tilde\phi T r)\right)\nonumber\\
\eea
Here $\tilde\phi\equiv \phi-\pi$ and $\mp$ refers to $M,\bar M$ respectively.
 Asymptotically, the BPS zero modes take the compact form in the hedgehog gauge

\bea
\label{ZERO2}
&&(\tilde\psi_{M} \epsilon )(r)\rightarrow \frac{1+{\rm  sgn}(\tilde\phi) \sigma \cdot \hat r}{\sqrt{2\pi T \nu r\sinh(2\pi T \nu r)}}
e^{|\tilde\phi|Tr}\nonumber\\
&&(\tilde\psi_{\bar M} \epsilon )(r)\rightarrow \frac{1-{\rm sgn}(\tilde\phi) \sigma \cdot \hat r}{\sqrt{2\pi T \nu r\sinh(2\pi T \nu r)}}
e^{|\tilde\phi|Tr}
\eea
For the KK instanton-dyon, we recall the additional time-dependent gauge transformation from the BPS 
instanton-dyon. The explicit form of the zero modes are also similar (\ref{ZERO1}-\ref{ZERO2}) with now
$\tilde \phi=\phi$. We note that for the flavor twisted boundary condition (\ref{BOUND1}), $f=d,s$ corresponds to $\tilde\phi=\mp \pi/3$ 
(mod $2\pi$) in (\ref{ZERO2}) which are not normalizable BPS zero modes at  exactly $\nu=1/3$.  
Following our analysis in~\cite{LIU5}, we choose to regulate the
zero modes by approaching the holonomies in the center symmetric phase as follows
($\epsilon_{1,2}\rightarrow +0$)

\bea
\label{CENTER3}
\nu_{M1}=&&\frac{1}{3}+\epsilon_1\nonumber\\
\nu_{M2}=&&\frac{1}{3}-\epsilon_2\nonumber\\
\nu_L=&&\frac{1}{3}+\epsilon_2-\epsilon_1
\eea
As a result, the M1-instanton-dyon carries 2 zero modes (d,s), the M2-instanton-dyon carries none, and the
L-dyon carries 1 zero mode (u). This regularization enforces the Nye-Singer index theorem for fundamental quarks~\cite{NS}
and the discrete symmetry noted earlier.

\subsection{Case: $N_c=N_f=2$}

For the case of $N_f=N_c=2$, a more general set of twisted boundary conditions will be analyzed with

\bea
\label{BOUND4}
&&u(\beta)=e^{i\theta}(-u(0))\nonumber\\
&&d(\beta)=e^{i\theta}(-e^{i\pi}d(0))
\eea
which is (\ref{BOUND1}) for $\theta=0$. (\ref{BOUND4}) is seen to have the additional discrete
symmetry  when $\theta \rightarrow \pi-\theta$ and $u\leftrightarrow d$ at $\nu=1/2$. Thus, only
the range $\theta<\pi/2$ will be considered. In this case, the M-instanton-dyon 
carries 1 zero-mode (d), while the L-instanton-dyon carries 1 zero-mode (u).
For (\ref{BOUND4}) the normalizable zero modes are asymptotically  of the form (\ref{ZERO2}) 
with $\phi=\theta$.

For completeness we note the Roberge-Weiss boundary condition~\cite{RW}

\bea
\label{BOUND5}
&&u(\beta)=e^{i\theta}u(0)\nonumber\\
&&d(\beta)=e^{i\theta}d(0)
\eea
In the range $0<\theta <\pi/2$, the M-instanton-dyon carries 2 zero modes with none on the L-instanton-dyon.
In the range $\frac{\pi}{2}<\theta<\frac{3\pi}{2}$, the 2 zero modes jump onto the L-instanton-dyon. 
In the range $0<\frac{3\pi}{2}<\theta<2\pi$  they jump back on the M-instanton-dyon. 
We note that for $\theta=\theta_0+\pi/2$ with $0<\theta_0<\pi/2$,
the M-zero mode moves to be an L-zero mode with the asymptotic

\be
\label{BXX1}
\frac{(1-\sigma \cdot \hat r)}{\sqrt{r\sinh(\pi T r)}}e^{(\pi/2-\theta_0)Tr}e^{i(\theta_0-\pi/2)T x_4}e^{i\pi Tx_4}
\ee
This is to be compared to the case with $\theta=\frac{\pi}{2}-\theta_0$ with the asymptotic

\be
\label{BXX2}
\frac{(1+\sigma \cdot \hat r)}{\sqrt{r\sinh(\pi T r)}}e^{(\pi/2-\theta_0)Tr}e^{i(\frac{\pi}{2}-\theta_0)Tx_4}
\ee

\subsection{Twisted fermionic determinant}

The fermionic determinant 
can be viewed as a sum of closed fermionic loops connecting all instanton-dyons and instanton-antidyons. Each link 
-- or hopping -- between an instanton-dyon  and ${\rm \bar{L}}$-anti-instanton-dyon is described by the hopping chiral
matrix

\begin{eqnarray}
\label{T12}
\tilde {\bf T}(x,y)\equiv \left(\begin{array}{cc}
0&i{\bf T}_{ij}\\
i{\bf T}_{ji}&0
\end{array}\right)
\end{eqnarray}
Each of the entries in ${\bf T}_{ij}$ is a  ``hopping amplitude" of a fermionic
zero-mode $\varphi_D$ from an  instanton-dyon to a  zero-mode
$\varphi_{\bar D}$ (of opposite chirality) of an instanton-anti-dyon

\be
{\bf T}_{LR}(x_{LR})=\int d^4x \varphi_{L}^{\dagger}(x-x_L)i(\partial_{4}-i\sigma\cdot\nabla )\varphi_R(x-x_R)\nonumber\\
{\bf T}_{RL}(x_{LR})=\int d^4x \varphi_{R}^{\dagger}(x-x_L)i(\partial_{4}+i\sigma\cdot\nabla )\varphi_L(x-x_R)\nonumber\\
\ee
with $x_{LR}\equiv x_L-x_R$, 
and similarly for the other components.  In the hedgehog gauge, these matrix elements can be made
explicit in momentum space. Their Fourier transform is

\be
\label{TPX}
T_{LR} (p)={\rm Tr}\left(\varphi_L^{\dagger}(p)(-\Phi T-i\sigma\cdot p)\varphi_R(p)\right)
\ee
with $\Phi T$ the contribution  from the lowest 
Matubara mode retained. We recall that the use of the zero-modes in the string gauge 
to assess the hopping matrix elements, introduces only minor changes in the overall estimates
as we discussed in~\cite{LIU2} (see Appendix A).

\subsubsection{Case $N_c=N_f=3$}

For general $\nu$, we use  the Fourier transform of the  zero modes
(\ref{ZERO1}) in (\ref{TPX}) to obtain 

\bea
\label{TPXX}
T_{i}(p)=\Phi_i T(F_{2i}^2(p)-F_{1i}^2(p))+{\rm sgn}(\tilde\phi_i)2pF_{1i}(p)F_{2i}(p)\nonumber\\
\eea
The key physics in the Fourier transforms $F_{1,2}(p)$ is captured  by retaining only the flux-induced 
mass-like  in the otherwise massless asymptotics, i.e. 

\be
F_{1i}(p)\approx\frac 13 F_{2i}(p)\approx \frac{\omega_0}{(p^2+((\nu-|\tilde\phi_i| /\pi)\omega_0)^2)^{\frac 54}}
\ee
 The i-assignments are respectively given by

\be
\label{ASSIGN}
i\equiv  (\bar L L, \bar M_1M_1, \bar M_2M_2) \qquad 
\biggl\{^{\tilde\phi_i=\left(0, -\frac {\pi}3, +\frac {\pi}3\right)}_{\Phi_i=(\pi,-\frac {\pi}3, +\frac {\pi}3)}\biggr.
\ee
In the center symmetric phase with $\nu=1/3$, (\ref{TPXX}) are long-ranged for the M-instanton-dyons,

\be
\label{HOP3}
T_3(p)=-T_2(p)\approx \Phi T\frac{8C^2}{p^5}+{\rm sgn}(\tilde\phi)\frac{6C^2}{p^4}
\ee
Here $C$ is a normalization constant fixed by the regularization detailed in (\ref{CENTER3}).

\subsubsection{Case $N_c=N_f=2$}

For $N_c=N_f=2$,  the 
Fourier transform of the lowest Matsubara zero-mode  for both boundaries (\ref{BOUND4}-\ref{BOUND5}) is

\be
\psi_M(p)=f_1(p)-i{\rm sgn} (\theta) f_2(p)\sigma\cdot \hat p
\ee
The correponding hopping matrix is ($0\leq \theta<\pi/2$)

\be
\label{HOPSU2}
T_{LR}(p)=\tilde \theta T(f_2^2(p)-f^2_1(p))+{\rm sgn}(\theta) 2pf_1(p)f_2(p)
\ee
with the assignments

\be
\tilde\theta=\biggl\{^{\theta-\pi\,\,: u}_{\theta\,\,\,\,\,\,\,\,\,\,: d}\biggr.
\ee
and

\bea
\label{F1F2}
&&f_1(p)\approx \frac 13 f_2(p)\approx \frac{\omega_0}{(p^2+((\nu_i-\theta /\pi)\omega_0)^2)^{\frac 54}} 
\eea
It follows that

\be
\label{ASSIGNX}
T_{LR}(p)\approx f_1(p)^2(8\tilde \theta T+6\, {\rm sgn}(\theta)\, p)
\ee
Using (\ref{BXX1}-\ref{BXX2}) we note that the hopping matrix
element  (\ref{ASSIGNX}) satisfies the anti-periodicity condition 

\be
\label{ANTIX}
T_{LR}(p,\theta_0+\pi/2)=-T_{LR}(p,\theta_0-\pi/2)
\ee
with the $\theta$-argument exhibited for clarity.

\section{SU($N_c$) ensemble}

Following~\cite{DP,LIU1,LIU2} the moduli
determinants in (\ref{SU2}) can be fermionized using $2N_c$ pairs of ghost fields $\chi_{m},\chi^{\dagger}_{m}$ for the 
instanton-dyons
and $2N_c$ for the instanton-anti-dyons. The ensuing Coulomb factors from the determinants are then bosonized using $2N_c$ boson fields $v_m,w_m$ for the instanton-dyons and similarly for
the instanton-anti-dyons.  The result is 

\bea
&&S_{1F}[\chi,v,w]=-\frac {T}{4\pi}\int d^3x\nonumber\\
&&\sum_{m=1}^{N_c}\left(| \nabla \chi_m|^2+\nabla v_m \cdot \nabla w_m\right)+\nonumber\\
&&\sum_{\bar m=1}^{N_c}\left(| \nabla \chi_{\bar m}|^2+\nabla v_{\bar m} \cdot \nabla w_{\bar m}\right)
\label{FREE1}
\eea
For the streamline interaction part $V_{D\bar D}$, we note that as a pair 
interaction in (\ref{SU2}) between the instanton-dyons and instanton-anti-dyons, it  can be bosonized using
standard methods~\cite{POLYAKOV,KACIR}  in terms of $\vec \sigma$ and $\vec b$ fields.   As a result each dyon species acquire additional fugacity factors of the form

\be
M:e^{-\vec\alpha_i \cdot \vec b+i\vec \alpha_{i}\cdot \vec \sigma} \qquad  \bar M:e^{-\vec\alpha_i \cdot \vec b-i\vec \alpha_{i}\cdot \vec \sigma}
 \ee
 with $\vec\alpha_i$ and $i=1,2, ...N_c-1$ the ith root of the $SU(N_c)$ Lie generator, and $i=N_c$ its affine root
 due to its compacteness. Therefore, there is an additional contribution to the free part (\ref{FREE1})

\be
S_{2F}[\sigma, b]=\frac T{8} \int d^3x\, \left(\nabla \vec b\cdot\nabla \vec b+ \nabla \vec \sigma\cdot\nabla\vec \sigma\right)
\label{FREE2}
\ee
where for simplicity we approximated the streamline by a Coulomb interaction, and the interaction part is now

\bea
&&S_I[v,w,b,\sigma,\chi]=-\int d^3x \nonumber\\
&&\left(\sum_{i=1}^{N_c}e^{-\vec\alpha_i \cdot \vec b+i\vec \alpha_{i}\cdot \vec \sigma}f_i\right.\nonumber\\
&&\times\left. \left(4\pi v_i+|\chi_i    -\chi_{i+1}|^2+v_i-v_{i+1}\right)e^{w_i-w_{i+1}}\right.\nonumber\\
&&\left.+\sum_{\bar i=1}^{N_c}e^{-\vec\alpha_{\bar i} \cdot \vec b-i\vec \alpha_{\bar i}\cdot \vec \sigma}f_{\bar i}\right.\nonumber\\
&&\left.\times  \left(4\pi v_i+|\chi_{\bar i}    -\chi_{\bar i+1}|^2+v_{\bar i}-v_{\bar i+1}\right)e^{w_{\bar i}-w_{\bar i+1}}\right)\nonumber\\
\label{FREE3}
\eea
without the fermions. We now show the minimal modifications to (\ref{FREE3}) when the fermionic determinantal
interaction is included.

\subsection{Fermionic fields}

To fermionize the determinant in (\ref{SU2})
and for simplicity, consider first the case of $N_f=1$  fermionic zero-modes attached to the kth instanton-dyon, and
define the additional Grassmanians $\chi=(\chi^i_1,\chi^j_2)^T$ with $i,j=1,.., K_{k,\bar k}$ so that

\be
\left|{\rm det}\,\tilde{\bf T}\right| =\int   D[\chi]\,\, e^{\,\chi^\dagger \tilde {\bf T} \, \chi}
\label{TDET}
\ee
We can re-arrange the exponent in (\ref{TDET}) by defining  a Grassmanian source $J(x)=(J_1(x),J_2(x))^T$ with

\be
J_1(x)=\sum^{K_L}_{i=1}\chi^i_1\delta^3(x-x_{ki})\nonumber\\
J_2(x)=\sum^{K_{\bar L}}_{j=1}\chi^j_2\delta^3(x-y_{\bar k j})
\label{JJ}
\ee
and by introducing 2 additional fermionic fields  $ \psi_k(x)=(\psi_{k1}(x),\psi_{k2}(x))^T$. Thus

\be
e^{\,\chi^\dagger \tilde {\bf T}\,\chi}=\frac{\int D[\psi]\,{\rm exp}\,(-\int\psi_k^\dagger \tilde {\bf G}\, \psi_k +
\int J^\dagger \psi_k + \int\psi_k^\dagger J)}{\int d
D[\psi]\, {\rm exp}\,(-\int \psi_k^\dagger \tilde {\bf G} \,\psi_k) }
\label{REFERMIONIZE}
\ee
with $\tilde{\bf G}$ a $2\times 2$ chiral block matrix

\begin{eqnarray}
 \tilde {\bf G}= \left(\begin{array}{cc}
0&-i{\bf G}(x,y)\\
-i{\bf G}(x,y)&0
\end{array}\right)
\label{GG}
\end{eqnarray}
with entries ${\bf TG}={\bf 1}$. The Grassmanian source contributions in (\ref{REFERMIONIZE}) generates a string
of independent exponents for the L-instanton-dyons and $\bar{\rm L}$-instanton-anti-dyons

\begin{eqnarray}
\prod^{K_k}_{i=1}e^{\chi_1^i\dagger \psi_{k1}(x_{ki})+\psi_{k1}^\dagger(x_{ki})\chi_1^i}\nonumber \\ \times
\prod^{K_{\bar k}}_{j=1}e^{\chi_2^j\dagger \psi_{k2}(y_{\bar k j})+\psi_{k2}^\dagger(y_{\bar k j})\chi_2^j}
\label{FACTOR}
\end{eqnarray}
The Grassmanian integration over the $\chi_i$ in each factor in (\ref{FACTOR}) is now readily done to yield

\be
\prod_{i}[-\psi_{k1}^\dagger\psi_{k1}(x_{ki})]\prod_j[-\psi_{k2}^\dagger\psi_{k2}(y_{\bar k j})]
\label{PLPR}
\ee
for the k-instanton-dyon and $\bar {\rm k}$-instanton-anti-dyon.
The net effect of the additional fermionic determinant in (\ref{SU2}) is to shift the k-instanton-dyon
and $\bar{\rm k}$-instanton-anti-dyon fugacities in (\ref{FREE3}) as follows

\bea
f_k\rightarrow -f_k\psi_{k1}^\dagger\psi_{k1}\equiv -f_L\psi_k^\dagger\gamma_+\psi_k\nonumber\\
f_{\bar k}\rightarrow -f_{\bar k}\psi_{k2}^\dagger\psi_{k2}\equiv -f_{\bar k}\psi_k^\dagger\gamma_-\psi_k
\label{SUB}
\eea
where we have now identified the chiralities with $\gamma_\pm=(1\pm \gamma_5)/2$. Note that
for the instanton-dyons and instanton-anti-dyons   with no zero-mode attached, the fugacities remain unchanged.

\subsection{Resolving the constraints}

In terms of (\ref{FREE1}-\ref{FREE3})  and the substitution
(\ref{SUB}), the instanton-dyon partition function (\ref{SU2})
for finite $N_f$ can be exactly re-written as an interacting
effective field theory in 3-dimensions,

\bea
{\cal Z}_{1}[T]\equiv &&\int D[\psi]\,D[\chi]\,D[v]\,D[w]\,D[\sigma]\,D[b]\,\nonumber\\&&\times
e^{-S_{1F}-S_{2F}-S_{I}-S_\psi}
\label{ZDDEFF}
\eea
with the additional chiral fermionic contribution $S_\psi=\psi^\dagger\tilde{\bf G}\,\psi$.
Since the effective action in (\ref{ZDDEFF}) is linear in the $v_{M,L,\bar M,\bar L}$, the latters
integrate to give the following constraints

\bea
\label{DELTAX}
&&-\frac{T}{4\pi}\nabla^2w_k+f_ke^{
\vec\alpha_{k}\cdot (-\vec b+i
\vec \sigma)}\prod_f \psi_{kf}^\dagger\gamma_+\psi_{kf}  e^{w_k-w_{k+1}}\nonumber\\&&
-f_{k-1}e^{\vec\alpha_{k-1}\cdot (-\vec b+i
\vec\sigma)}\prod_f \psi_{k-1f}^\dagger\gamma_+\psi_{k-1f}\,\e^{w_{k-1}-w_k}=0\nonumber\\
\eea
and similarly for the anti-dyons.

To proceed further the formal classical solutions to the constraint equations or $w[\sigma, b]$
should be inserted back into the 3-dimensional effective action. The result is

\bea
{\cal Z}_{1}[T]=\int D[\psi]\,D[\sigma]\,D[b]\,e^{-S}
\label{ZDDEFF1}
\eea
with the 3-dimensional effective action

\bea
&&S=S_F[\sigma, b]+\int d^3x\,\sum_f \psi_f^\dagger \tilde{\bf G}_f \psi_f\nonumber\\
&& +\sum_{k=1}^{N_c}4\pi f_kv_k\int d^3x\,\prod_{f} \psi_{kf}^\dagger\gamma_+\psi_{kf}\,e^{w_k-w_{k+1}+\vec\alpha_{k}\cdot (-\vec b+\vec i\sigma)}\nonumber\\
&&+\sum_{\bar k=1}^{N_c}4\pi f_{\bar k}v_{\bar k}\int d^3x\,\prod_{f} \psi_{\bar kf}^\dagger\gamma_-\psi_{\bar kf}
\,e^{w_{\bar k}-w_{\bar k+1}+\vec\alpha_{\bar k}\cdot (-\vec b+\vec i\sigma)}\nonumber\\
\label{NEWS}
\eea
Here $S_F$ is $S_{2F}$ in (\ref{FREE2}) plus additional contributions resulting from the $w(\sigma, b)$ solutions
to the constraint equations (\ref{DELTAX}) after their insertion back.  This procedure for the linearized approximation of the constraint
was discussed in~\cite{LIU1,LIU2}.

For the general case with

\be
\tilde{\bf G}_1\neq \tilde{\bf G}_2\neq ...\neq \tilde{\bf G}_{N_f}
\ee
these contributions in (\ref{NEWS})  are only
$U_L^{N_f}(1)\times U_R^{N_f}(1)$ symmetric, which is commensurate with (\ref{SYMNF}). The determinantal interactions preserve
the individual $U_{L+R}(1_k)$ vector flavor symmetries, but upset the individual $U_{L-R}(1_k)$ 
axial flavor symmetries. However, the latters induce the shifts

\be
\label{BACK0}
\psi_{kf}^\dagger\gamma_\pm \psi_{kf}\rightarrow e^{2\xi_k}\psi_{kf}^\dagger\gamma_\pm \psi_{kf}
\ee
which can be re-absorbed by shifting back the constant magnetic contributions

\be
\label{BACK}
\vec\alpha_{\bar k}\cdot (-\vec b+\vec i\sigma)\rightarrow \vec\alpha_{\bar k}\cdot (-\vec b+\vec i\sigma)-2\xi_k
\ee
thanks to the free form in (\ref{FREE2}).  This observation  is unaffected by the screening of the 
magnetic-like field, since a constant shift $\vec b\rightarrow \vec b+2\xi_k$ can always be reset
by a field redefinition. This hidden symmetry was noted recently in~\cite{THOMAS}.
We note that this observation holds  for the general form of the 
streamline interaction used in~\cite{LIU2} as well, due to its vanishing form in momentum space. From
(\ref{BACK}) it follows that $\sum_k\xi_k=0$, so that only the axial flavor singlet $U_{L-R}(1)$ is explicitly
broken by the determinantal contributions in (\ref{NEWS}) as expected in the instanton-dyon-anti-dyon
ensemble. As a result, (\ref{NEWS}) is  explicitly $U(1)_L^{N_f}\times U_R^{N_f}(1)/U_{L-R}(1)$ symmetric.

\subsection{Special cases: $N_c=N_f=2,3$}

For the case $N_c=N_f=3$ with the twisted  boundary condition (\ref{BNC}), the fermionic
terms in the effective action (\ref{NEWS}) are explicitly

\bea
\label{SF3X}
&&\psi^{\dagger}_u \tilde G_1\psi_u+\psi_d^{\dagger}\tilde G_2\psi_d+\psi_s^{\dagger}\tilde G_3\psi_s\nonumber \\
&&+4\pi f_1\nu_1\psi^{\dagger}_u\gamma_{+}\psi_ue^{w_1-w_2}\nonumber\\
&&+4\pi f_2\nu_2\psi_d^{\dagger}\gamma_{+}\psi_d\psi_s^{\dagger }\gamma_{+}\psi_se^{w_2-w_3}
+4\pi f_3\nu_3e^{w_3-w_1}\nonumber\\
&&+4\pi f_{\bar 1}\bar\nu_{1}\psi^{\dagger}_u\gamma_{-}\psi_ue^{\bar w_1-\bar w_2}\nonumber\\
&&+4\pi f_{\bar 2}\bar\nu_{2}\psi_d^{\dagger}\gamma_{-}\psi_d\psi_s^{\dagger }\gamma_{-}\psi_se^{\bar w_2-\bar w_3}
+4\pi f_{\bar 3}\bar\nu_{3}e^{\bar w_3-\bar w_1}\nonumber\\
\eea
following the regulartization (\ref{CENTER3}) around the center symmetric point. As noted earlier, 
(\ref{SF3X}) is explicitly symmetric under the combined
charge conjugation and the flavor exchange $d\leftrightarrow s$ since $\tilde G_2=-\tilde G_3\neq \tilde G_1$. 
With this in mind, (\ref{SF3X}) is symmetric under
$(U^3_L(1)\times U^3_R(1))/U_{L-R}(1)$.

For the case $N_c=N_f=2$ with the twisted boundary condition (\ref{BOUND4}), the fermionic
terms in the effective action (\ref{NEWS}) are now

\bea
&&f_Mv_M\psi_d^{\dagger}\gamma_{+}\psi_de^{w_M-w_L}+f_Lv_L\psi_u^{\dagger}\gamma_{+}\psi_ue^{w_L-w_M} \nonumber\\
&&+f_{\bar M}v_{\bar M}\psi_d^{\dagger}\gamma_{-}\psi_de^{w_{\bar M}-w_{\bar L}}
+f_{\bar L}v_{\bar L}\psi_u^{\dagger}\gamma_{-}\psi_ue^{ w_{\bar  L}-w_{\bar M}} \nonumber\\
\eea
while for the Roberge-Weiss boundary condition (\ref{BOUND5}) they are

\bea
&&f_Mv_M\psi_u^{\dagger}\gamma_{+}\psi_u\psi_d^{\dagger}\gamma_{+}\psi_de^{w_M-w_L}+f_Lv_Le^{w_L-w_M} \nonumber\\
&&+f_ {\bar M}v_{\bar M}\psi_u^{\dagger}\gamma_{-}\psi_u\psi_d^{\dagger}\gamma_{-}\psi_de^{w_{\bar M}-w_{\bar L}}
+f_{\bar L}v_{\bar L}e^{w_{\bar L}-w_{\bar M}} \nonumber\\
\eea

\section{Equilibrium state}

To analyze the ground state and the fermionic fluctuations we  bosonize the fermions
in (\ref{ZDDEFF1}-\ref{NEWS})  by introducing the identities

\bea
\label{deltax}
&&\int D[\Sigma_k]\,\delta\left(\psi^\dagger_k(x)\psi_k(x)+2\Sigma_k(x)\right)={\bf 1}
\eea
and by re-exponentiating them to obtain

\bea
{\cal Z}_{1}[T]=\int D[\psi]\,D[\sigma]\,D[b]\,D[\vec\Sigma]\,D[\vec\lambda]\,
e^{-S-S_C}\nonumber\\
\label{ZDDEFF2}
\eea
with

\bea
\label{SCx}
&&-S_C=\int d^3x \,i\Lambda_k(x)(\psi_f^{\dagger}(x)\psi_k(x)+2\Sigma_k(x))
\eea
The ground state is parity even so that $f_{L,M}=f_{\bar L, \bar M}$.
By translational invariance, the  ground state corresponds to constant $\sigma, b, \vec\Sigma, \vec\Lambda$.
We will seek the extrema of (\ref{ZDDEFF2}) with finite condensates in the mean-field approximation, i.e.

\bea
\label{DEFCC}
&&\left<\psi^\dagger_k(x)\psi_l(x)\right>=-2\delta_{kl}\Sigma_k
\eea
With this in mind, the classical solutions to the constraint equations (\ref{DELTAX}) are also constant

\bea
\label{SU2SOL}
&&f_k \left< \prod _f\psi^{\dagger}_{kf}\gamma_{+}\psi_{kf} \right>e^{w_k-w_{k+1}}
\nonumber \\ 
&&=f_{k+1}\left<\prod_f \psi_{k+1f}^\dagger\gamma_+\psi_{k+1f}\right>\,\e^{w_{k+1}-w_{k+1}}
\eea
with

\bea
\label{SU2SOLx}
\left<\prod_f \psi_{kf}^\dagger\gamma_+\psi_{kf}\right>=\prod_{f}\Sigma_{kf}
\eea
and similarly for the anti-dyons. The expectation values in (\ref{SU2SOL}-\ref{SU2SOLx}) 
are carried in  (\ref{ZDDEFF2}) in the mean-field approximation through Wick contractions.

\subsection{$N_c=N_f=3$ in symmetric phase}

In the   center-symmetric phase, with all holonomies being equal $\nu_{1,2,3}=1/3$, the pressure simplifies to

\bea
\label{P33}
{\cal P}_{uds}-{\cal P}_{per}=&&8\pi (f_1f_2f_3)^{\frac 13}(\Sigma_{u}\Sigma_{d}\Sigma_{s})-2\vec \Lambda\cdot \vec \Sigma\nonumber \\
&&+\sum_{i=1}^{3}\int \frac{d^3p}{(2\pi)^3}\ln(1+\Lambda_i^2|T_{i}|^2(p))\nonumber\\
\eea
with the individual fermionic terms being
\bea
{\cal P}_i\equiv &&\int \frac{d^3p}{(2\pi)^3}\ln(1+\Lambda_i^2|T_{i}|^2(p))\nonumber\\
\equiv &&\omega_0^3\int  \frac{d^3 \tilde p}{(2\pi)^3}
\ln \left(1+\frac{\tilde \Lambda_i^2}{\tilde p^8}\left(1+\frac{4|\tilde\phi_i|}{3\pi\tilde p}\right)^2\right)
\eea
Here $\tilde p=p/\omega_0$ and $\tilde\Lambda_i=\Lambda/\omega^2_0$ are dimensionless. From (\ref{ASSIGN}),
we  recall  the  assignment of quark phases $(\tilde\phi_1,\tilde\phi_2,\tilde\phi_3)=(\pi,-\pi/3, +\pi/3)$, for $(u,d,s)$ respectively. 
The center symmetric phase breaks spontaneously chiral symmetry, as the gap equation have nonzero
solution. Each of the flavor chiral
condensate is found to be
\be
\label{QQ3}
\frac{\left<\bar q q\right>_{\tilde\phi_i}}{T^3}=2\pi^2\tilde \Lambda_i
\int \frac{d^3\tilde p}{(2\pi)^3}\frac{\frac{5}{3\tilde p^5}}{1+\frac{\tilde \Lambda_i^2}{\tilde p^8}\left(1+\frac{4|\tilde\phi_i|}{3\pi\tilde p}\right)^2}
\ee

We now note that at asymptotically low temperatures, the $1/p^4$ contribution in the hopping matrix element (\ref{HOP3}) is dominant.


%
%
%
%
%
%
%

\subsection{$N_c$=$N_f=3$ in general asymmetric phase}

In general asymmetric phase the holonomies have values away from the center 
\bea
\label{H123}
\nu_1=&&\frac{1}{3}+\epsilon_1\nonumber\\
\nu_2=&&\frac{1}{3}-\epsilon_2\nonumber\\
\nu_3=&&1-\nu_1-\nu_2
\eea
Note that in general,  the parameters $\epsilon_{1,2}$ are not small. 
With these choices for the holonomies (\ref{H123}) , the u-flavor rides the L-instanton-dyon, 
and the ds-flavors ride the $M1,M2$-instanton-dyons. For the ds-flavors, the hopping matrix elements between the 
instanton-dyon and anti-instanton-dyon are   given by

\bea
&&T_d(p)=-T_s(p)=\nonumber\\
&&\frac{\pi T}{3}(F_2^2(p)-F_1^2(p))+2ipF_1(p)F_2(p)
\eea
with 

\be
\label{XF1X}
F_1(p)\approx \frac 1{3}{F_2}(p)\approx \frac{\omega_0}{(p^2+((\nu_1-1/3)\omega_0)^2)^{\frac{5}{4}}}
\ee
while for the u-quarks it is

\be
T_u(p)=\pi T(f_2^2(p)-f_1^2(p))+2ipf_2(p)f_1(p)
\ee
with

\be
f_1(p)\approx \frac 1{3} {f_2}(p)\approx \frac{\omega_0}{(p^2+(\nu_3\omega_0)^2)^{\frac{5}{4}}}
\ee

In the mean-field approximation, the modification of the effective pressure  is

\bea
\label{Puds}
{\cal P}_{uds}-{\cal P}_{per}&&=+24\pi(f_1f_2f_3\nu_1\nu_2\nu_3\Sigma_d^2\Sigma_u)^{\frac 13}\nonumber\\
&&-4\Sigma_d\Lambda_d-2\Sigma_u\Lambda_u\nonumber\\
&&+\int \frac{d^3p}{(2\pi)^3}\ln\left((1+\Lambda_d^2|T_d|^2)^2(1+\Lambda_u^2|T_u|^2)\right)\nonumber\\
\eea
where ${\cal P}_{\rm per}$ is the perturbative contribution with twisted quark boundary conditions~\cite{RW}.
For $\nu_1\rightarrow 1/3$ the holonomy  induced mass-like  contribution in (\ref{XF1X})  becomes arbitrarily small.
As we noted earlier, we use it to regulate the infrared sensitivity of the ds-contributions in (\ref{Puds}) through a suitable
redefinition of the fugacities $f_{2,3}$ as in~\cite{LIU5}.  With this in mind, 
the extrema of (\ref{Puds}) with respect to $\Sigma, \Lambda$ yield the 
respective gap equations

\bea
\label{SADX}
\Lambda_d=&&4\pi f(\nu_1\nu_2\nu_3)^{\frac 13}\left(\frac{\Sigma_u}{\Sigma_d}\right)^{\frac 13}\nonumber\\
\Lambda_u=&&4\pi f(\nu_1\nu_2\nu_3)^{\frac 13}\left(\frac{\Sigma_d}{\Sigma_u}\right)^{\frac 23}\nonumber\\
\Sigma_i=&&\int  \frac{d^3p}{(2\pi)^3}\frac{\Lambda_i|T_i(p)|^2}{1+\Lambda_i^2|T_i(p)|^2}
\eea
Using (\ref{SADX}) in (\ref{Puds}) results in the shifted pressure at the saddle point

\bea
&&{\cal P}_{uds}-{\cal P}_{per}=\nonumber\\
&&\int \frac{d^3p}{(2\pi)^3}{\rm ln}
\left[(1+\Lambda_d^2|T_d|^2)^2 (1+(\frac{\tilde \Lambda_0^3}{\Lambda_d^2})^2|T_u|^2)\right]\nonumber\\
\eea
with $\tilde \Lambda_0=4\pi f(\nu_1\nu_2\nu_3)^{\frac 13}$.
We note that the gap equation follows from ${d{\cal P}}/{d\Lambda_d}=0$.
The  chiral condensates follow from standard arguments as 

\bea
\left<\bar d d\right>=\left<\bar s s\right>=&&2\Lambda_d T
\int \frac{d^3p}{(2\pi)^3}\frac{F_1^2(p)+F^2_2(p)}{1+\Lambda_d^2|T_{d}(p)|^2}\nonumber\\
\left<\bar u u\right>=&&2\Lambda_u T
\int \frac{d^3p}{(2\pi)^3}\frac{f_1^2(p)+f^2_2(p)}{1+\Lambda_u^2|T_{u}(p)|^2}
\eea

In contrast and at asymptotically high temperatures, the $1/p^5$ contribution in the hopping matrix element (\ref{HOP3}) is dominant.
Therefore the  u-hopping is different from the d- and s-hoppings with $T_{1}(p)\approx 3T_{2}(p)$. The extrema 
of the pressure in $\Lambda_{1,2,3}$ are now found to be

\be
3\Lambda_1=\Lambda_2=\Lambda_3=\frac{4\pi T} 3 (3\nu_1\nu_2\nu_3f_1f_2f_3)^{\frac 13}
\ee
with distinct chiral condensates

\be
\label{QQRATIO}
3\left<\bar u u\right>\approx \left<\bar d d\right>\approx \left<\bar s s\right>\approx 0.78\,T^3(\tilde \Lambda_2)^{\frac 35}
\ee
The high temperature phase breaks flavor symmetry but preserves the discrete combined charge conjugation symmetry
and the exchange $d\leftrightarrow s$. 
As a check on these observations, we note that for $\tilde \Lambda \approx 1$, the chiral condensates
in (\ref{QQ3}) are numerically close

\bea
\left<\bar q q\right>_{\tilde\phi=\pi} \approx && 0. 61T^3\nonumber\\
\left<\bar q q\right>_{\tilde\phi=\frac \pi 3}\approx && 0.76T^3
\eea

The remaining task is to solve the gap equations for the
four remaining parameters $\Lambda_d,\Lambda_u,\epsilon_1,\epsilon_2$.
The numerical analysis of those equations  will be presented elsewhere.

\subsection{$N_c=N_f=2$ in symmetric phase}

The analysis of the $N_f=N_c=2$  follows similar arguments using the
twisted boundary conditions (\ref{BOUND4}) for $\pi\nu>\theta$. In this case the 
u-flavor rides the L-dyon, and the d-flavor rides the M-dyon with the hopping
matrices

\bea
\label{TUPXX}
T_u(p)=&&(\pi -\theta )T({\tilde f}_2^2(p)-{\tilde f}_1^2(p))+2ip{\tilde f}_1(p){\tilde f}_2(p)\nonumber\\
T_d(p)=&&\theta T (f_2(p)^2-f_1^2(p))+2ipf_1(p)f_2(p)
\eea
with

\be
\label{f12X}
f_1(p)\approx \frac 13 f_2(p)\approx \frac{\omega_0}{(p^2+((\nu-\theta /\pi)\omega_0)^2)^{\frac 54}}
\ee
${\tilde f}_{1,2}$ follows from $f_{1,2}$ using the substitution $\theta\rightarrow -\pi+\theta$. 
We note that for $\theta=0$, the first contribution in $T_d$ vanishes, since the d-boundary is
periodic with zero Matsubara frequency. It is proportional to the Matsubara frequency in $T_u$,
since the u-boundary is anti-periodic. This difference is in addition to the different mass-like contributions
induced by the holonomy (d: $\nu\omega_0$ and u: $\tilde\nu\omega_0$), 
which regulate the small-momenta (large distance) behavior of the hopping amplitudes
and
causes the flavor condensates to be relatively different.

In the mean-field limit, the non-perturbative pressure  is

\bea
\label{P22}
{\cal P}_{ud}-{\cal P}_{\rm per}=&& 16\pi f(\nu_1\nu_2\Sigma_1\Sigma_2)^{\frac 12}-2\Lambda_1\Sigma_1-2\Lambda_2\Sigma_2\nonumber \\
&&+\sum_{i=1,2}\int \frac{d^3p}{(2\pi)^3}\ln(1+\Lambda_i^2|T_{i}(p)|^2)
\eea
while the perturbative one (with our twisted boundary conditions) is given by

\bea
&&{\cal P}_{\rm per}=-\frac{4\pi^2T^3}3\left(\nu_1\nu_2\right)^2\nonumber\\
&&-\frac{4T^3}{\pi^2}\sum_f\sum_{n=1}^{\infty}\frac{(-1)^ne^{i\theta_fn}}{n^4}{\rm Tr}_{f}L^n\nonumber\\
\eea
The first contribution comes from the gluons, while the second contribution  comes from
the twisted quarks. The Polyakov line $L$ is in the fundamental representation,
with the flavor twist explicitly factored out.
The dominant contribution in the sum stems from the $n=1$ term. Note that
for  $\theta_1=0$ and $\theta_2=\pi$, the fermionic contribution almost cancels.

The gap equations related to the parameters $\Lambda_i,\Sigma_i$ are
\bea
\label{GAP2F}
&&\Lambda_1=4\pi f(\nu_1\nu_2)^{\frac 12}\left(\frac{\Sigma_2}{\Sigma_1}\right)^{\frac{1}{2}}\nonumber\\
&&\Lambda_2=4\pi f(\nu_1\nu_2)^{\frac 12}\left(\frac{\Sigma_1}{\Sigma_2}\right)^{\frac{1}{2}}\nonumber\\
&&\Sigma_i=\int \frac{d^3 p}{(2\pi)^3}\frac{\Lambda_i|T_i|^2}{1+\Lambda_i^2|T_i(p)|^2}
\eea
The chiral condensates are readily obtained as

\be
\left<\bar q_i q_i\right>=2\Lambda_iT\int \frac{d^3p}{(2\pi)^3}\frac{f_1^2(p)+f^2_2(p)}{1+\Lambda_i^2|T_{i}(p)|^2}
\ee
We note that for large $\Lambda$ or asymptotically small temperatures,
 the second term in (\ref{ASSIGNX}) proportional to $p$ is dominant. In this case, the
 hopping matrix elements for $M,L$ are equal. It follows  that the extrema of
 the pressure (\ref{P22}) are also equal, 
 
\be
\Lambda\equiv \Lambda_{1}=\Lambda_{2}=2\pi (f_Lf_M)^{\frac 12}
\ee
In this limit, the chiral condensates are also the same

\be
\label{ASUD}
\left<\bar u u\right>\approx\left<\bar d d\right>\approx 2\Lambda T
\int \frac{d^3p}{(2\pi)^3}\frac{f_1^2(p)+f^2_2(p)}{1+\Lambda^2|T_{1,2}(p)|^2}
\ee
with $f_{1,2}(p)$ given in (\ref{F1F2}).


Before we discuss  the general asymmetric case, let us make the following 
comments on the so called Roberge-Weiss symmetry~\cite{RW}.
Since the hopping matrix elements satisfy the anti-periodicity 
condition (\ref{ANTIX}),  the pressure (\ref{P22}) satisfies the so called
$half$-periodicity condition

\be
{\cal P}(\theta+\pi/2)={\cal P}(\theta-\pi/2)
\ee
in the center symmetric phase. Using the explicit form (\ref{ASSIGNX}), we find that 

\be
\label{CUSPFREE}
\left(\frac{d{\cal P}}{d\theta}\right)_{\theta\rightarrow \pi/2}=0
\ee
which is cusp free despite the switching of the zero-mode from the M- to L-instanton-dyon.
These observations are in agreement with those put forth
by Roberge and Weiss~\cite{RW} at low temperatures. At high temperature 
(\ref{CUSPFREE}) develops a cusp in the center asymmetric phase~\cite{RW}.
We have checked that these properties hold also for the twisted boundary condition
(\ref{BOUND4}).

\subsection{$N_c=N_f=2$: general asymmetric case}

To proceed, we first note that the gap equations  (\ref{GAP2F})  can be simplified by noting that
$\Lambda_1\Lambda_2=n^2$ and that $\Lambda_2\Sigma_2=\Lambda_1\Sigma_1$.
We have set  $n=4\pi f(\nu_1\nu_2)^{\frac 12}$ with $\nu_1=\nu$ and $\nu_2=1-\nu$. 
With this in mind, (\ref{GAP2F}) reduces to a single gap equation,

\be
\label{GAPXXX0}
\int d^3\tilde p\,\frac{|\tilde T_1|^2}{1/\tilde\Lambda_1^2+|\tilde T_1|^2}=
\int d^3\tilde p\,\frac{|\tilde T_2|^2}{\tilde\Lambda_1^2/\tilde n^4+|\tilde T_2|^2}
\ee
After  rescaling of all variables $\tilde p=p/\omega_0$, $\tilde\Lambda_{1,2}=\Lambda_{1,2}/\omega_0^2$ 
and $\tilde n=n/\omega_0^2$ with $\omega_0=\pi T$,  the hopping matrices (\ref{TUPXX}) simplify
\bea
\label{GAPXXX}
|\tilde T_1|^2\approx &&\frac{(6\tilde p)^2}{(\tilde p^2+\nu_1^2)^5}\nonumber\\
|\tilde T_2|^2\approx &&\frac{64+(6\tilde p)^2}{(\tilde p^2+\nu_2^2)^5}
\eea
 After using the gap equations (\ref{GAP2F}) and the rescaling, the pressure (\ref{P22}) becomes

\bea
\label{PREY}
\frac{{\cal P}_{ud}}{\omega_0^3}=&&\int \frac{d^3\tilde p}{(2\pi)^3}
{\rm ln}\left[(1+\tilde\Lambda_1^2|\tilde T_1|^2)(1+(\frac{\tilde n^2}{\tilde\Lambda_1})^2|\tilde T_2|^2)\right]\nonumber\\
&&-\frac{4\pi^2}3\frac {T^3}{\omega_0^3}(\nu_1\nu_2)^2
\eea
Its extremum in $\Lambda$ is the gap equation  $\partial{\cal P}_{ud}/\partial\tilde\Lambda_1=0$, which  is (\ref{GAPXXX0}).
Similarly, there is the gap equation for the holonomy $\nu$. The task is to solve them together. 
We  found that (\ref{PREY}) leads to the  
momentum-dependent constituent masses
for the d-, u-quarks 
\bea
\label{MASSud}
&&\frac{M_d(p)}{\omega_0}\equiv (1+\tilde p^2)^{\frac 12}\tilde\Lambda_1|\tilde T_1(p)|\nonumber\\
&&\frac{M_u(p)}{\omega_0}\equiv (1+\tilde p^2)^{\frac 12}\frac{\tilde n^2}{\tilde\Lambda_1}|\tilde T_2(p)|
\eea
The u-quark is subtantially heavier than the d-quark at low momentum because of its 
anti-periodic boundary condition, with the d-quark  turning 
massless at zero momentum owing to its periodic boundary condition. 

\begin{figure}[h!]
 \begin{center}
 \includegraphics[width=8cm]{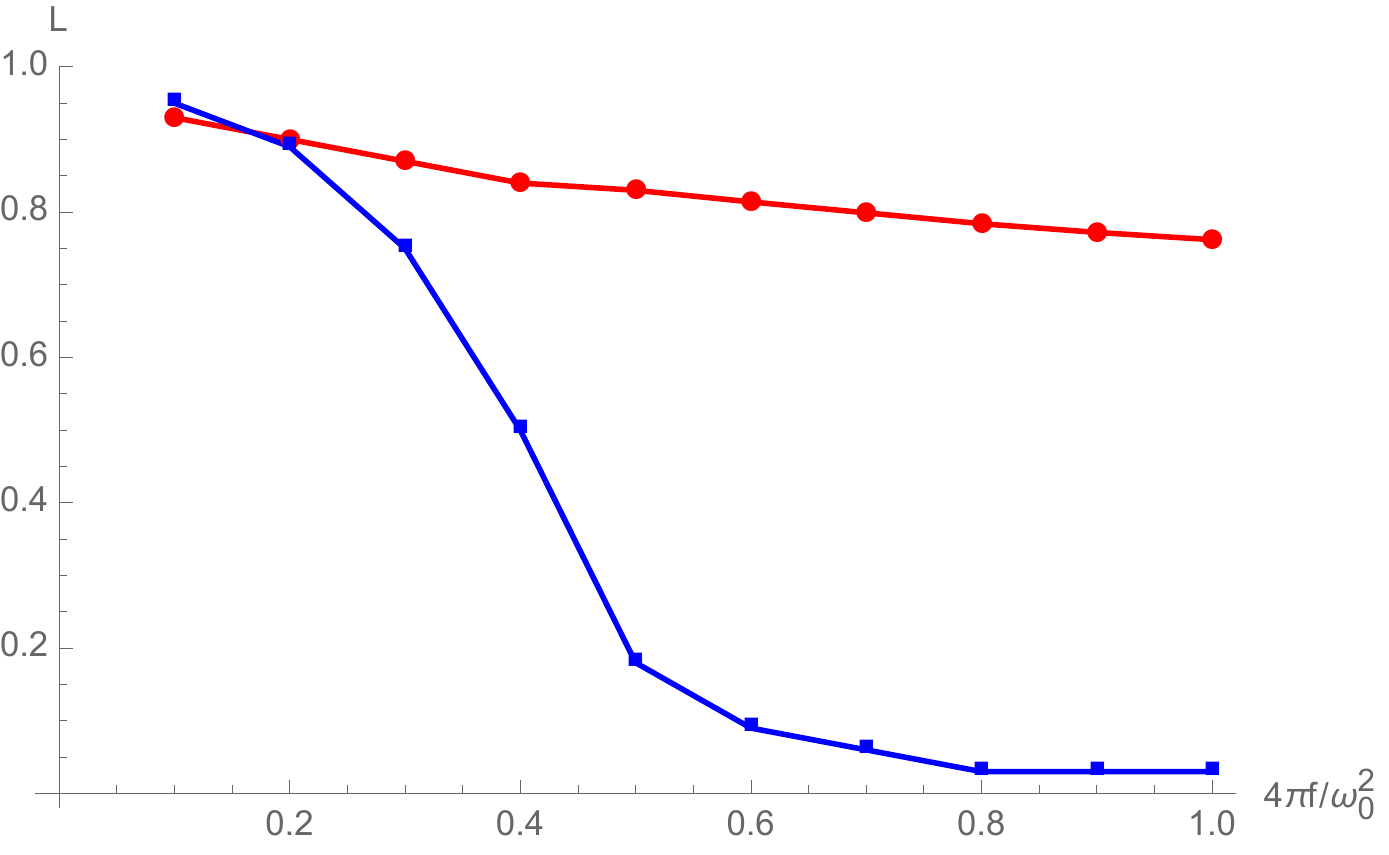}
 \caption{ Polyakov line versus the dimensionless 
 density ${\bf n}=4\pi f/\omega_0^2$ for $N_f=N_c=2$. The lower (square-blue) line is for the $Z_2$ twisted 
 quarks, while the upper (circle-red) line is for the usual anti-periodic quarks.}
 \label{fig_ll}
  \end{center}
\end{figure}

\begin{figure}[h!]
 \begin{center}
 \includegraphics[width=8cm]{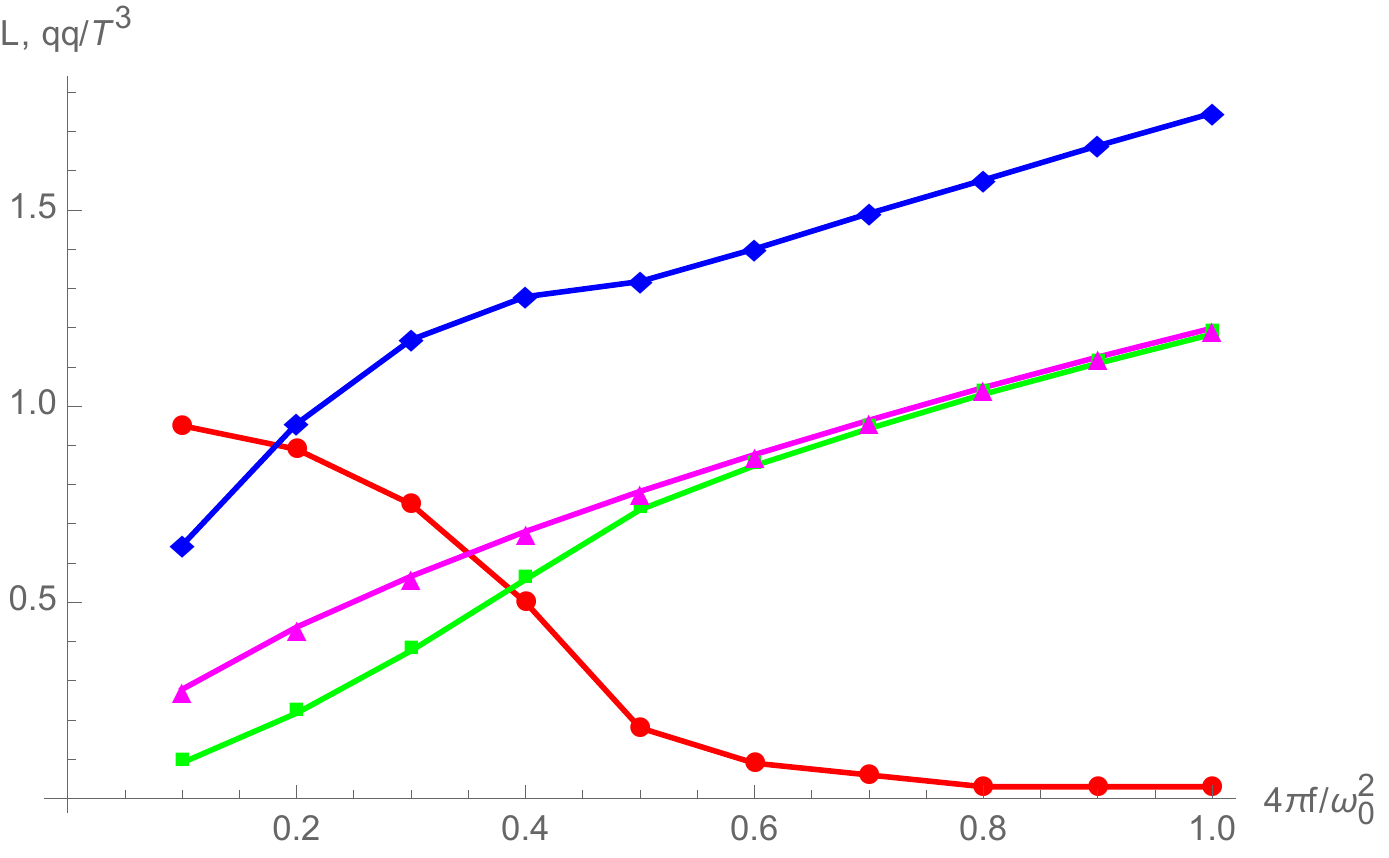}
 \caption{Dimensionless condensates  $|\left<\bar d d\right>|/T^3$ (diamond-blue), 
  $|\left<\bar uu \right>|/T^3$ (square green) for twisted boundary conditions, 
  with increasing dimensionless density or lower temperatures $4\pi f/\omega_0^2$.
 For comparison we show  $|\left<\bar uu\right>|/T^3$  (triangle-magenta) for the anti-periodic quarks. The Polyakov line (square-red)
 shows a rapid crossing from a center broken to a center symmetric phase for the twisted quarks. }
   \label{fig_lud}
  \end{center}
\end{figure}

The results for the  numerical solution of the gap equations are shown in Fig.~\ref{fig_ll} 
and Fig.~\ref{fig_lud} . In Fig.~\ref{fig_ll}, we show the dependence of the Polyakov
line $L={\rm cos}(\pi \nu[{\bf n}])$ on the input parameter ${\bf n}=4\pi f/\omega_0^2$ (square-blue) in the lower line. For
comparison we also show the behavior of the same Polyakov line (circle-red) in the upper  line, 
for the untwisted (QCD) theory with both u-, d-quarks being anti-periodic fermions.
The input parameter ${\bf n}$  is a definite monotonously decreasing function of the temperature as defined in (\ref{NTT}).
 The rightmost part of the plot corresponds to the dense low-$T$ case, in which we find a confining or $L\rightarrow 0$ behaviour.
 The main conclusion from this plot is that confinement (or restoration of center symmetry) occurs at a lower 
density ${\bf n}$ for the twisted theory, as compared to the QCD-like one.

In Fig.~\ref{fig_lud} we show the behavior of the flavor condensates $|\left<\bar d d\right>|/T^3$ (upper-diamond-blue),
$|\left<\bar uu\right>|/T^3$ (lower-square-green) for the twisted u-, d-quarks versus ${\bf n}=4\pi f/\omega_0^2$. 
For comparison, we also show the value of $|\left<\bar uu \right>|/T^3=|\left<\bar d d\right>|/T^3$ (middle-triangle-magenta)
for the untwisted  (anti-periodic) bounday conditions. It  follows closely the line for the anti-periodic d-quark in the twisted case. 
The value of the Polyakov line for the twisted quarks is shown 
also (circle-red), to indicate the transition region. At high densities or low temperatures, center symmetry is restored but the quark condensates are still distinct for
the twisted boundary condition.  The induced effective masses in (\ref{MASSud}) show that the d-quark is much lighter than
the u-quark, resulting in a much larger chiral condensate.  Only at vanishingly small temperatures, the relation (\ref{ASUD})  is recovered as both hoppings become identical. The  nature of the  boundary condition becomes irrelevant at zero temperature. 
At low densities or high temperatures,  center symmetry is broken and the  chiral condensate
$|\left<\bar d d\right>|$ is still substantially larger than  $|\left<\bar uu\right>|$.

\subsection{Mesonic spectrum}

The excitation spectrum  with twisted boundary conditions can be calculated
following the analysis in~\cite{LIU2}. For the $N_c=N_f=2$ case, this follows by
substituting

\bea
\label{MES1}
\Lambda(\psi^{\dagger}\gamma_{\pm}\psi+2\Sigma^{\pm})\rightarrow \sum_{fg}\Lambda^{\pm}_{fg}(\psi^{\dagger}_f\gamma_{\pm}\psi_g+2\Sigma^{\pm}_{fg})
\eea
in (\ref{SCx}) with 

\be
\Lambda_{\pm}\equiv \Lambda_0\pm i\pi_{ps}+\pi_s
={\rm diag}(\Lambda_1,\Lambda_2)\pm i\pi_{ps}+\pi_s
\ee
Here $\pi_{s,ps}$ refer to the scalar and pseudo-scalar $U(2)$-valued mesonic fields. 

For the chargeless chiral partners  $\sigma^3, \pi^0$, the effective actions to quadratic order 
are respectively given by

\bea
\label{MES2}
S(\pi_{ps}^3)=&&\frac{1}{2f_{\pi}^2}\int \frac{d^3p}{(2\pi)^3}\pi_{ps}^3(p)\Delta_{-}^3(p)\pi_{ps}^3(-p)\nonumber\\
S(\pi_{s}^3)=&&\frac{1}{2f_{\pi}^2}\int \frac{d^3p}{(2\pi)^3}\pi_{s}^3(p)\Delta_{+}^3(p)\pi_{s}^3(-p)
\eea
with the corresponding propagators ($p_\pm =q\pm p/2$)

\bea
\label{MES3}
\Delta_{\pm}^3(p)=\frac{1}{2}\int \frac{d^3q}{(2\pi)^3}\frac{(T_1(p_+)\pm T_1(p_-))^2}{(1+\Lambda_1^2|T_1(p_+)|^2)(1+\Lambda_1^2|T_1(p_-)|^2)}\nonumber \\+\frac{1}{2}\int \frac{d^3q}{(2\pi)^3}\frac{(T_2(p_+)\pm T_2(p_-))^2}{(1+\Lambda_2^2|T_2(p_+)|^2)(1+\Lambda_2^2|T_2(p_-)|^2)}\nonumber\\
\eea
with the hopping matrices $T_{1,2}$  labeled as $1\equiv d$ and $2\equiv u$. 
In deriving (\ref{MES2}-\ref{MES3})  we made explicit use  of the gap equations (\ref{GAP2F}). We note that $\Delta^3_-(0)=0$
translates to a massless $\pi^0=\pi^3_{ps}$, while $\Delta^3_+(0)\neq 0$ translates to a massive $\sigma$,
for both the center symmetric and broken phases. The masslessness of $\pi^0$ is ensured by the hidden symmetry displayed in 
(\ref{BACK0}-\ref{BACK}), and reflects on the remaining spontaneously broken symmetry for $N_f=2$.

The charged mesons  $\pi_{s}^{\pm},\pi_{ps}^{\pm}$, follow a similar analysis with now the propagators
for the quadratic contributions given by 

\be
\Delta_{\pm}^{1,2}(p)=\frac{(\Sigma_1\Sigma_2)^{\frac 12}}{\pi f}-2\int \frac{d^3q}{(2\pi)^3}\,{\mathbb F}_\mp (p,q)
\ee
Here  $\Delta_-^{1,2}$ refer to the charged scalars $\pi_s^\pm$,  while $\Delta_+^{1,2}$ refer to their
charged chiral partners  $\pi_{ps}^\pm$, with

\be
{\mathbb F}_\pm (p,q)=\frac{T_{1}(p_+)T_2(p_-)(\Lambda_1\Lambda_2T_{1}(p_+)T_2(p_-)\pm 1)}{(1+\Lambda_1^2|T_1(p_+)|^2)(1+\Lambda_2^2|T_2(p_-)|^2)}
\ee
In the exactly center symmetric  phase, with $\Lambda_1=\Lambda_2$, the charged pions $\pi_{ps}^\pm$ 
are also massless. But in general, in the  asymmetric phase $\Lambda_1\neq \Lambda_2$, and
both $\pi^\pm$ are massive (but degenerate).

The singlet mesons $\sigma=\pi_{s0},\eta=\pi_{ps,0}$  propagators follow similarly

\bea
&&2\Delta_{\sigma}(p)=\frac{n_D}{2}+\Delta_{+}^3(p)\nonumber\\
&&2\Delta_{\eta}(p)=\frac{n_D}{2}+\Delta_{-}^3(p)
\eea
with $n_D$ the mean instanton-dyon density defined through the gap equation
\be
\frac{n_D}{4}=\frac{1}{2}\sum_{i=1}^2\int \frac{d^3p}{(2\pi)^3}\frac{\Lambda_i^2|T_i|^2}{1+\Lambda_i^2|T_i|^2}
\ee

\section{Conclusions}

We have constructed  the partition function for the instanton-dyon liquid model with twisted flavor
boundary conditions, and derived and solved the resulting gap equations in the mean-field approximation. 
In addition to manifest $U^{N_F}(1)\times U^{N_f}(1))/U_{L-R}(1)$
flavor symmetries, for $Z_{N_c}$-QCD some  discrete charge conjugation plus
flavor exchange symmetries were identified .

The central constructs are the so called hopping matrix elements between 
instanton-dyon and anti-instanton-dyon zero modes. One technical point is
to note that some of these hoppings may become singular at large distances 
(small momenta) when the contribution from the $Z_{N_c}$-twists and the
holonomies cancel the exponentially decreasing asymptotics. 
These singularities are readily  regulated through a suitable redefinition
of the pertinent fugacities~\cite{LIU5}.

 The low temperature phase is  center symmetric with zero Polyakov line. It also breaks chiral symmetry, with 
 still sizably different chiral condensates in our mean-field analysis. The latters are about equal at very small
 temperatures.  The high temperature phase is center asymmetric with  always
 unequal chiral condensates. Our results are qualitatively consistent with the 
 lattice results reported recently in~\cite{TAKUMI}, although with a more pronounced difference between the
 flavor chiral condensates across the transition region caused mostly by the differences in the leading (twisted) Matsubara modes
 in the center symmetric phase.  In the symmetric ground state we observe the emergence of one massless pion $\pi^0$
 (2-flavor case).

 The instanton-dyon model offers a very concise framework for
 discussing the interplay of twisted boundary 
 conditions (also known as flavor holonomies) with 
 center symmetry and chiral symmetry in the QCD-like models. 
A further comparison between the mean field results derived in this paper, with the direct
simulations \cite{LARS3} of the instanton-dyon model and lattice results \cite{TAKUMI},
is obviously of great interest.

\section{Acknowledgements}

We thank Takumi Iritani for an early discussion.
This work was supported by the U.S. Department of Energy under Contract No.
DE-FG-88ER40388.

 \vfil

\end{document}